\documentclass[twocolumn,superscriptaddress,pra,letterpaper,showpacs]{revtex4}
\usepackage{amsmath}
\usepackage{graphicx,color}
\usepackage{float}
\usepackage[caption=false]{subfig}
\usepackage{verbatim} 
\usepackage{enumerate}
\usepackage{blkarray}
\usepackage{amsfonts}
\usepackage{amssymb}
\usepackage{hyperref}

\newcommand{\ket}[1]{\left\vert#1\right\rangle}

\newcommand{\beq}{\begin{equation}}
\newcommand{\eeq}{\end{equation}}
\newcommand{\baq}{\begin{eqnarray}}
\newcommand{\eaq}{\end{eqnarray}}
\newcommand{\unit}{1\!\!1}
\newcommand{\brac}[1]{\lbrace #1\rbrace}

\newcommand{\p}{\mathtt{p}}

\begin{document}

\title{Optimal digital dynamical decoupling for general decoherence via
        Walsh modulation}
        
\author{Haoyu Qi} 
\affiliation{Hearne Institute for Theoretical
        Physics and Department of Physics \& Astronomy,Louisiana State
        University, Baton Rouge, Louisiana 70803, USA} 

\author{Jonathan P. Dowling} 
\affiliation{Hearne Institute for Theoretical
        Physics and Department of Physics \& Astronomy,Louisiana State
        University, Baton Rouge, Louisiana 70803, USA} 

\author{Lorenza Viola}
\affiliation{\mbox{Department of Physics \& Astronomy, Dartmouth College, 6127 Wilder Laboratory, 
                Hanover, New Hampshire 03755, USA}}

\begin{abstract}
We provide a general framework for constructing digital dynamical decoupling sequences 
based on Walsh modulation --- applicable to arbitrary qubit decoherence scenarios. 
By establishing equivalence between 
decoupling design based on Walsh functions and on concatenated projections,
we identify a family of {\em optimal Walsh sequences}, which can be exponentially more efficient, 
in terms of the required total pulse number, for fixed cancellation order, than known digital sequences 
based on concatenated design. Optimal sequences for a given cancellation order are highly non-unique 
--- their performance depending sensitively on the control path. 
We provide an analytic upper bound to the achievable decoupling error, and 
show how sequences within the optimal Walsh family can substantially outperform 
concatenated decoupling, while respecting realistic timing constraints.
We validate these conclusions by numerically computing the average fidelity in a toy model capturing 
the essential feature of hyperfine-induced decoherence in a quantum dot.
\end{abstract}
\pacs{03.65.Yz,03.67.Pp,89.70.+c}
\date{\today}
\maketitle

Dynamical decoupling (DD) techniques, based on open-loop quantum control, provide an effective 
strategy to reduce decoherence from temporally correlated 
noise processes in realistic quantum information processing platforms \cite{dd,qec}.
In its simplest form, DD coherently averages out the unwanted system-environment interaction through 
the application of tailored sequences of (ideally, instantaneous) pulses, whose net action on the system 
translates, in the frequency domain, into a high-pass noise filter \cite{Kofman:04,Green:njp,fff,soare2014}.  
To date, the most efficient DD schemes known for generic error models 
--- notably, Uhrig DD \cite{uhrig2007keeping} and quadratic DD \cite{west2010near} for 
pure dephasing and general decoherence on a single qubit 
--- involve pulse sequences with irrational pulse timing \cite{remarknjp}.  However, 
consideration of practical constraints highlights crucial advantages of {\em digital} DD, 
whereby all pulse separations are integer multiples of an experimentally restricted 
minimum time interval. Irrationally-timed DD sequences have been found
to be more sensitive to both the form of the spectral cutoff and to 
inevitable pulse errors \cite{ryan2010robust,xiao2011efficiency,ajoy2011optimal}, while being less 
amenable to the additional compensation steps (e.g., via phase-shifts or composite pulses) that are 
needed to mitigate these errors for arbitrary input states 
\cite{souza2012robust,wang2012comparison,farfurnik2015optimizing,genov2016}. 
Even in situations where pulse imperfections are unimportant, 
digital DD sequences are highly compatible with hardware constraints stemming from 
digital sequencing circuitry and clocking, which makes them attractive in terms of 
minimizing sequencing complexity, as ultimately demanded for large-scale 
implementations. 
 
Control modulation based on Walsh functions \cite{beauchamp1975walsh}, has 
been proposed as a unifying approach for generating { digital-efficient} protocols, for 
both dynamically corrected quantum storage and gates 
\cite{hayes2011reducing,hayes2012,HarrisonWalsh,violaMemory}.
Walsh DD (WDD) has been shown to naturally incorporate existing digital sequences 
as special instances (including concatenated DD for both single- and multi-axis decoherence 
\cite{cdd}), and to provide a restricted search space for numerical sequence optimization and 
analytic performance analysis under finite timing resources. 
For dephasing noise on a qubit, concatenated DD sequences based on single-axis control 
are provably optimal, among the Walsh suite, in the sense of guaranteeing 
a desired order of error suppression with minimum total pulse number \cite{hayes2011reducing}.

In this work, we identify optimal single-qubit WDD sequences capable of 
canceling out the simultaneous dephasing and relaxation 
effects that arise from arbitrary environmental couplings.  
The key step is to generalize existing sequence constructions of WDD based on multi-axis control, 
and establish formal equivalence of the resulting general WDD formalism with the 
{concatenated-projection DD} (CPDD) approach proposed in Ref. \cite{qi2015method}.
By leveraging this equivalence, we explicitly characterize the error-suppression 
capabilities of {\em any} general WDD sequence, 
along with its complexity in terms of the required control time slots.
We show that, unlike in the dephasing scenario, concatenated DD is no longer optimal, 
and identify a large family of {\em optimal} WDD (OWDD) schemes, whose
complexity is exponentially smaller for the same order of suppression.
While the performance of different OWDD sequences depends additionally 
on the specific control path, our analysis indicates that OWDD 
can substantially improve over existing digital schemes in relevant parameter regimes.
 
\textit{Control-theoretic setting.}---We consider a single-qubit system $S$ coupled to an 
uncontrollable quantum environment (bath) $B$ 
via an arbitrary interaction, that is, we let the joint evolution in the absence of control to 
be generated by a Hamiltonian of the form 
$H \equiv H_{S}\otimes \unit_B + H_{SB} + \unit_S\otimes H_B$, with 
$H_{SB} = \sigma_x\otimes B_x + \sigma_y\otimes B_y + \sigma_z\otimes B_z$.
Here, $H_{S}$ and $H_B\equiv B_0$ are, respectively, the internal Hamiltonian for $S$ and $B$ alone, 
and $\sigma_u$, $u\in \{x,y,z\}$, denote qubit Pauli matrices.
The bath operators $B_0, B_u$ are assumed to be bounded but otherwise arbitrary (possibly unknown). 
In what follows, we shall use $\beta \equiv \vert \vert B_0||$ and $J \equiv \max_{u\in \{x,y,z\}} \brac{||B_u||}$ to 
quantify the strength of the internal-bath dynamics vs. the system-bath interaction, with $|| \cdot||$ being the operator norm.

DD is implemented via a control action on $S$ alone, generated by a control Hamiltonian 
of the form $H_{c}(t)\otimes \unit_B$. In this work, we consider digital DD sequences consisting of 
ideal instantaneous $\pi$-pulses along any of the three coordinate axes.  Thus, for a sequence 
specified by pulse timings and control operations $\{t_j,\sigma_j \}$, involving a total of $N$  
pulses over a running time $T$, the control Hamiltonian  
$H_{c}(t) = \frac{\pi}{2} \sum_{j=1}^N \sigma_j\delta(t-t_j)$, 
where we let $t_0\equiv 0, t_N\equiv T$, and $\sigma_j \in \{\sigma_u\}, \forall j$.
Crucially, the digital constraint mandates that all inter-pulse separations obey 
$t_j - t_{j-1} \equiv n_j \tau_0$, with $n_j \in \mathbb{N}$, with   
the minimum pulse interval $\tau_0 >0$ determined by hardware limitations. 
Another convenient representation we shall use for the above sequence is 
$P_N\mathtt{f}_{n_N\tau_0}\ldots P_2\mathtt{f}_{n_2\tau_0}P_1\mathtt{f}_{n_1\tau_0}$, 
where the $P_j \in \brac{X,Y,Z}$ represent different $\pi$ pulses and $\mathtt{f}_{n_j\tau_0}$ 
denotes free evolution between $P_{j-1}$ and $P_{j}$. 

Since the DD objective is to achieve an identity gate on $S$, all the evolution 
induced by $H$ contributes to unwanted error dynamics \cite{dcgs}, whereby $H \equiv H_e$. 
Let  $U_c(t) \equiv \mathcal{T}\exp[-i\int_0^t H_{c}(t')dt']$ be the control propagator, 
with $\hbar =1$ and $\mathcal{T}$ denoting time-ordering. 
The effect of $H_e$ may be isolated by expressing the propagator $U(T)$, for evolution 
under $H(t) \equiv H_e +H_c(t)$ over time $T$, as 
$U(T) = U_c (T) \,
\mathcal{T}\exp[-i\int_0^t \tilde{H}_e(t')dt'] \equiv e^{-i \Omega_e(T)}$, 
where $U_c(T)=\unit_S$ for DD. $\tilde{H}_e(t) = U_c^\dagger(t) H_e U_c(t)$ describes 
evolution in the toggling frame associated to $H_c(t)$, and    
$\Omega_e(T)$ defines the error action operator \cite{dcgs}. 
The norm of $\Omega_e(T)$, up to pure-bath terms that do not enter the reduced 
dynamics, quantifies the achievable error per gate (EPG). 
Specifically, $\Omega_e(T)$ and the associate effective Hamiltonian may be obtained 
via a perturbative Magnus expansion, 
$ \Omega_e(T) \equiv [H_{SB}^{\text{eff}} (T) + H_{B}^{\text{eff}} (T) ]T = 
\exp[\sum_{m=1}^\infty \Omega_e^{(m)}(T)]$, where $\Omega_e^{(m)}(T)$ is a 
time-ordered integral involving $m^{\text{th}}$-order nested commutators, and 
$||H || T< \pi$ suffices for (absolute) convergence \cite{Blanes:09}.  
The DD performance in the time domain is then 
characterized by the order of error suppression ({cancellation order}, CO),  
determined by the leading correction mixing $S$ and $B$ in $\Omega_e(T)$ 
\cite{dd,fff}.  That is, $\text{EPG} \equiv ||\text{mod}_B( \Omega_e(T))|| = 
|| T H_{SB}^{\text{eff}}(T) || = {\mathcal O}(T^{\alpha+1})$
for a protocol with $\text{CO} =\alpha$.

\vspace*{1mm}


\textit{Walsh vs. concatenated-projection DD formalism.}---The Walsh functions are a well-known family of 
binary-valued piecewise-constant functions orthonormal over $[0,1]$,  
which may be naturally employed to describe digital DD sequences \cite{beauchamp1975walsh,hayes2011reducing}. 
For dephasing noise, single-axis control via $\pi$-pulses around (say) the $x$-axis suffices in the 
ideal case, resulting in a control propagator of the form 
$U_c(t) \equiv \sigma_x^{[x(t)+1]/2}$,
where the control switching function $x(t)$ toggles between the values $\pm 1$ at instants 
corresponding to the applied pulse timings. Let the Walsh function of Paley order $n$
be defined as
\begin{eqnarray*}
W_n(x) \equiv \prod_{j=1}^{m}R_j(x)^{b_j},\quad x\in[0,1], 
\end{eqnarray*}
where $\brac{b_j}$ is the binary representation of $n$, namely $n = \sum_{j=1}^m b_j2^{j-1}$, 
and $R_j(x) \equiv \text{sgn}[\sin(2^j\pi x)]$ is the {Rademacher function}, 
which switches between $\pm 1$ with frequency $2^{j-1}$. 
A {\em WDD$_n$ sequence} is then defined as the pulse sequence with switching function
$x(t) = -W_n(t/T)$, $t\in[0,T].$
If $r \equiv \sum_{m}b_m$ is the Hamming weight of $n$ (hence the number of 
Rademacher functions used to construct $W_n(x)$), 
the corresponding WDD$_n$ protocol achieves $\text{CO}=r$ \cite{hayes2011reducing} . 

For a single qubit exposed to multi-axis decoherence, Ref.~\cite{hayes2011reducing} also defines 
two-axis WDD protocols by allowing for the control propagator $U_c(t)$ to involve 
{\em two} switching functions, say, for $\pi$-pulses along the $x$ and $y$ directions, with the form 
$x(t) = R_{j_1}R_{j_3}\ldots R_{j_{2r-1}}$, $y(t)= R_{j_2}R_{j_4}\ldots R_{j_{2r}}$. In this way, for 
$n=4^r -1$, the resulting WDD$_n$ protocol reproduces concatenated DD (CDD) of level $r$, 
again achieving $\text{CO}=r$ for this general error model \cite{cdd}.

A different approach to digital DD design is provided by CPDD \cite{qi2015method}, whereby 
pulse sequences are built by concatenating {projection sequences}. There are four such 
sequences, $\mathtt{p}_0 \equiv I\mathtt{f}_{\tau_0}I\mathtt{f}_{\tau_0}$, 
$\mathtt{p}_x \equiv X\mathtt{f}_{\tau_0}X\mathtt{f}_{\tau_0}$, and similarly for $\mathtt{p}_y$ 
and $\mathtt{p}_z$. Applying $\mathtt{p}_u$, with $u\in\brac{x,y,z}$, suppresses the interaction 
along perpendicular directions, to the first order,
that is, with corresponding $\text{EPG}=\mathcal{O}(\tau_0^2 ||H||^2)$ \cite{projection}.
Given two pulse sequences $A$ and $B$, their concatenation may be defined as 
$A[B] \equiv P^A_{N_A}(B)\dots P^A_2(B)P^A_1$. The new pulse sequence constructed in 
this way inherits the suppression capabilities from each of the original pulse sequences. 
Concatenating a pulse sequence with $\mathtt{p_0}$ corresponds to simply repeating the sequence twice. 
A {\em CPDD$_s$ sequence} is then specified by an ordered string $s \equiv 
s_m s_{m-1}\ldots s_1$, with $s_j \in\brac{0,x,y,z}$, 
with each symbol labeling a projection sequence. To construct the corresponding pulse sequence, 
projection sequences are concatenated according to the specified string, namely, 
$\text{CPDD}_s \equiv \p_{s_1}[\ldots[\p_{s_{m-1}}[\p_{s_m}]]].$
For example, in this notation CDD$_r = \text{CPDD}_{(xy)^r}$.


\textit{General WDD.}---Our first result is a generalization of multi-axis WDD 
beyond the existing one.  Unlike the construction in \cite{hayes2011reducing}, 
we start by expressing the control propagator in terms of {\em three} distinct 
switching functions: 
\begin{eqnarray}
U_c(t) = \sigma_x^{[x(t)+1]/2}\sigma_y^{[y(t)+1]/2}\sigma_z^{[z(t)+1]/2}.
\label{propagator}
\end{eqnarray} 
We define general WDD (GWDD) sequences as follows:

\vspace*{1mm}

{\em Definition.} {A GWDD$_{\vec{n}}$ sequence is specified by an integer vector 
consisting of three Paley orders, $\vec{n} \equiv (n_x,n_y,n_z),$ 
subject to the constraint
$\sum_{u=x,y,z}b^{u}_j \leq 1$, $1\leq j\leq m_u$. Here, $b^u_j$ is the $j$th digit 
in the binary representation of $n_u$, where $n_u =  \sum_{j=1}^{m_u} b_j^u 2^{j-1}$.
The switching function for control along direction $u$ in Eq. (\ref{propagator}) is 
\begin{eqnarray}
u(t) = -\text{WDD}_{n_u}(t/T) = \prod_{j=1}^{m} R_j^u (t/T)^{b_j^u}, \; 
t \in [0,T],
\end{eqnarray}
with $m\equiv \max\brac{m_x,m_y,m_z}$ and $b_j^u \equiv 0$ for $m_u<j \leq m$.}

\vspace*{1mm}

Since any $\pi$-pulse can be obtained as the product of two $\pi$-pulses along 
orthogonal directions, the constraint on the coefficients $b_j^u$ is necessary to avoid redundant 
sequences, by allowing {at most one non-zero digit} among all three digits at 
each binary location.  Clearly, the above definition recovers the one in Refs. 
\cite{hayes2011reducing,HarrisonWalsh},  
where a single integer suffices to specify a two-axis WDD$_n$, due to 
the assumed particular structure.
For instance, $r^{\text{th}}$-order CDD corresponds to a GWDD$_{\vec{n}}$,  
with $\vec{n} =  (2(4^r - 1)/(4-1),(4^r - 1)/(4-1),0)$, with the 
single above-mentioned Paley order $n=4^{r}-1$ 
being the sum of three Paley orders in our definition.


Crucially, the above GWDD definition is instrumental to both establish equivalence with the 
CPDD formalism, and uncover optimal GWDD sequences not accounted for otherwise. 
To demonstrate the equivalence, note that each non-zero digit $b_j^u$, in the binary representation 
of $n_u$ in a GWDD sequence, may be associated to a projection 
$\mathtt{p}_u$ in the equivalent CPDD sequence. When $b_j^u=0$ for all $u\in\brac{x,y,z}$, 
we have an identity projection $\mathtt{p}_0$ in CPDD. 
Explicitly, the following conversion rules hold:

\vspace*{1mm}

(i) {\em CPDD-to-GWDD.} Given a CPDD$_s$ with $s = s_ms_{m-1}\ldots s_1$, calculate $n_u = \sum_{j=1}^m b^u_j 2^{j-1}$ for $u \in\brac{x,y,z}$, where $b_j^u = 1$ if $s_j = u$, otherwise $b_j^\mu = 0$. The corresponding GWDD sequence is GWDD$_{n_x,n_y,n_z}$.

\vspace*{1mm}

(ii) {\em GWDD-to-CPDD.} Given a GWDD$_{n_x,n_y,n_z}$, first convert each Paley order to its binary representation,  $n_u = (b_{m_u}^\mu b_{m_{u}-1}^\mu\ldots b_1^u)_2$. Second, leftpad the binary representations with zeros so that they all have the same length $m$.
For the $j$th digit and $u\in\brac{x,y,z}$, set $s_j = u$ if $b_j^{u} = 1$; else, if all $b_j^u = 0$, set 
$s_j = 0$. The corresponding CPDD sequence is CPDD$_s$ with $s = s_ms_{m-1}\ldots s_1$.

\vspace*{1mm}

The resulting correspondence is illustrated in 
Table~\ref{tb:1} for single-axis sequences. For multi-axis control, we use the so-called 
GA$_{8r}$ sequences as an example. The latter is obtained from concatenation 
of a basic six-pulse, $2^{\text{nd}}$-order GA$_8$ sequence, 
$I \mathtt{f}X\mathtt{f} Y \mathtt{f}X \mathtt{f}I \mathtt{f}X \mathtt{f}Y \mathtt{f}X\mathtt{f}$, 
found by a genetic search algorithm in  Ref.~\cite{quiroz2013optimized}.
In the CPDD framework, GA$_{8r}$ = CPDD$_{(zyx)^r}$. By using the above rules, we have 
$n_x = (100\ldots 100)_2$, $n_y = (010\ldots 010)_2$, $n_z = (001\ldots 001)_2$. Accordingly, the 
corresponding GWDD sequence is GWDD$_{\vec{n}}$, where $\vec{n}$ is given by
$\vec{n} = (4(1-2^{3r})/(1-2^3), 2(1-2^{3r})/(1-2^3), (1-2^{3r})/(1-2^3))$.

\begin{table}[t]
        \begin{tabular}{c|c}
                \hline\hline
                WDD$_n$ & CPDD$_s$ \\ 
                \hline\hline
                WDD$_0$ & CPDD$_0$ = p$_0$\\
                \hline
                WDD$_1$ & CPDD$_x$ = p$_x$\\
                \hline
                WDD$_2$ = WDD$_{10}$ & CPDD$_{x0}$ = p$_0$[p$_x$]\\
                \hline
                WDD$_3$ = WDD$_{11}$ & CPDD$_{xx}$ = p$_x$[p$_x$]\\
                \hline
                WDD$_4$ = WDD$_{100}$ & CPDD$_{x00}$ = p$_0$[p$_0$[p$_x$]]\\
                \hline\hline
        \end{tabular}
        \caption{Equivalence between single-axis WDD and CPDD.}
        \label{tb:1}
\end{table}

The equivalence with the CPDD formalism makes it possible to easily obtain 
the CO of an arbitrary GWDD sequence. As shown in Ref.~\cite{qi2015method}, the CO of CPDD 
is given by $\alpha = \min\brac{r_y + r_z, r_x + r_z, r_x + r_y},$
where $r_u$ is the number of projections along the $u$-axis. From the above rules, we see that each such projection 
implies a non-zero bit in the binary representation of the corresponding Paley order. Therefore, the CO of GWDD is still given by 
the above equation, but with $\{ r_u=\sum_j b_j^u\}$ now being Hamming weights. 
It follows that GWDD/CPDD sequences with the same CO are highly 
non-unique: permuting the order of projections will produce a different GWDD 
sequence, but leave the CO unchanged. Accordingly, we may think of GWDD sequences specified by 
$(r_{{\cal P}(x)}, r_{{\cal P}(y)}, r_{{\cal P}(z)})$, where ${\cal P}\in {\cal S}_3$ is any permutation, as 
forming an equivalence class with respect to CO. 

\begin{table}[t]
	\label{multiprogram}
	\begin{tabular}{c|c|c|c||c|c|c}
		\hline
		& \multicolumn{3}{c||}{OWDD} & \multicolumn{3}{c}{CDD} \\
		\hline\hline
		CO & ($n_x,n_y,n_z$) & $N_T$ & $N$ & ($n_x,n_y,n_z$) & $N_T$ & $N$\\
		\hline
		1  & (2,1,0) & 4 & 4 & (2,1,0) & 4 & 4\\
		\hline
		2  & (4,2,1) & 8 & 6 & (10,5,0) & 16 & 14\\
		\hline
		3  & (18,5,4) & 32 & 32 & (42,21,0) & 64 & 60\\
		\hline
		4  & (36,18,9) & 64 & 42 & (170,85,0) & 256 & 238\\
		\hline\hline
	\end{tabular}
	\caption{Number of control time slots and applied pulses for OWDD vs. CDD.  
		Taking either $N_T$ or $N$ does not change the optimality of OWDD, as their differences 
		are negligible compared to the exponential saving.}
	\label{tab:pulse number}
\end{table}

\textit{Optimal Walsh DD.}---In the presence of a realistic constraint, $\tau_0>0$, 
achieving higher CO comes at the price of either increasing the total number of pulses $N$
for fixed storage time $T$ --- until the maximum CO compatible with the constraints is accommodated;
or of increasing both $N$ and $T$ --- until perturbative error suppression breaks down and, again,
a maximum CO is reached beyond which no further improvement occur \cite{violaMemory}.
This motivates defining  {\em optimal WDD} (OWDD) sequences by demanding that 
they guarantee a desired CO with minimum pulse number or, equivalently,   
{minimum number of time slots, $N_T$}, each slot having duration $\tau_0$.  
Within single-axis WDD, CDD sequences are provably optimal
\cite{hayes2011reducing}. However, this is no longer true  
for multi-axis GWDD. The optimal GWDD can be inferred from 
the CPDD framework. For $\text{CO}=\alpha$, let $\overline{\alpha} =\pm 1$ 
denote the parity of $\alpha$. Then, all GWDD$_{\vec{n}}$ satisfying the following two conditions 
are optimal and define an equivalence class referred to as OWDD$_\alpha$:
\begin{align}
\label{eq:OWDD-conditions-1}
&\sum\limits_{u = x,y,z} b_j^u = 1, \quad 1\leq j \leq m,\\
\label{eq:OWDD-conditions-2}
&(r_{{\cal P}(x)}, r_{{\cal P}(y)}, r_{{\cal P}(z)})= \frac{1}{2}\left(\alpha - \overline{\alpha},\alpha + \overline{\alpha}, \alpha + \overline{\alpha} \right).
\end{align}
Eq. (\ref{eq:OWDD-conditions-1}) ensures that the pulse sequence does not 
expend any pulse on repetition, whilst Eq. (\ref{eq:OWDD-conditions-2}) gives the Hamming weights 
(number of projections) needed to suppress decoherence up to the required CO. 
From the equivalence between CPDD and GWDD, and the analysis in 
Ref.~\cite{qi2015method}, it follows 
that OWDD$_\alpha$ uses a number of time slots given by
$\log_2(N_T) = \frac{1}{2}\left( 3\alpha+ \overline{\alpha}  \right).$
A comparison between OWDD$_\alpha$ and CDD$_\alpha$ is included in Table \ref{tab:pulse number}. 
If the CO is sufficiently large, OWDD$_\alpha$ is {\em exponentially more efficient} than CDD, since 
$ N_T^{\text{OWDD}_\alpha }/N_T^{\text{CDD}_\alpha} \approx 2^{\frac{3}{2}\alpha}/2^{2\alpha}
= 2^{-\alpha/2}.$


\textit{Performance analysis.}---The EPG provides an appropriate performance measure 
for control since it upper-bounds the trace-norm distance between the intended 
and the actual final states of the system, say, $\Delta(\rho_S^0(T), \rho_S(T))$, where 
$\rho_S^0(T)= \rho_S^0(0)\equiv |\psi \rangle \langle\psi |$ for DD \cite{lidar2008distance,dcgs}. 
That is, $\Delta( \rho_S^0(T), \rho_S(T) ) \leq || T H_{SB}^{\text{eff}} (T) ||$
independently of $\ket{\psi}$, which in turn allows us to bound experimentally accessible fidelities
as $ 1-\Delta \leq F \leq \sqrt{1-\Delta^2}$
(here, $F(\rho, |\psi \rangle \langle\psi |) 
\equiv \text{tr} \sqrt{ \sqrt{\rho} \,|\psi \rangle \langle\psi | \sqrt{\rho}}\,$). 
An analytical upper bound to the EPG 
may be derived by leveraging the geometrical picture 
afforded by CPDD. The basic idea is to observe that, if the error action operator for a 
sequence CPDD$_{s_0}$ with running time $T_{s_0}$ has the form $\Omega_e(T_{s_0}) \equiv 
T_{s_0} [ \sum_u \sigma_u \otimes B_u^{s_0} + B_0^{s_0}]$, 
concatenation with a projection sequence, say,  
$\p_x[\text{CPDD}_{s_0}] \equiv \text{CPDD}_{s_0 x}$, 
has a simple effect in terms of renormalizing 
bath operators in the orthogonal directions. Specifically, 
one 
finds \cite{supplement}
that the resulting error action, $\Omega_e(2 T_{s_0})$, 
has the same structure as $\Omega_e(T_{s_0})$, only with new bath operators:   
\vspace{-0mm}
\begin{subequations}
	\label{eq:norm_bound}
	\begin{align}	
	||B_0^{s_0 x}|| &=||B_0^{s_0}||, \quad   
	||B_x^{s_0 x}|| = ||B_x^{s_0}||, \\
	||B_y^{s_0 x}|| &\leq T_{s_0}(\beta||B_y^{s_0}|| + ||B_z^{s_0}||\, ||B_x^{s_0}||) ,\\
	||B_z^{s_0 x}|| &\leq T_{s_0}(\beta||B_z^{s_0}|| + ||B_y^{s_0}||\, ||B_x^{s_0}||) ,
	\end{align}
\end{subequations}
as long as $\beta \gg J$ or $\ll J$, and $\max\brac{\beta,J}\tau_0\ll 1$. Similar inequalities hold when we 
concatenate CPDD$_s$ with $\p_y$ or $\p_z$. 
Therefore, given an arbitrary GWDD sequence, the desired upper bound may be obtained by 
first translating it into the equivalent CPDD$_s$, and then by repeatedly using 
Eqs. (\ref{eq:norm_bound}) according to the ``projection path'' specified by $s= s_m \ldots s_1$, 
leading to $\text{EPG} \leq T_s \sum_{u=x,y,z}||B_u^s||$.

\begin{figure}[t]
\includegraphics[scale=0.5]{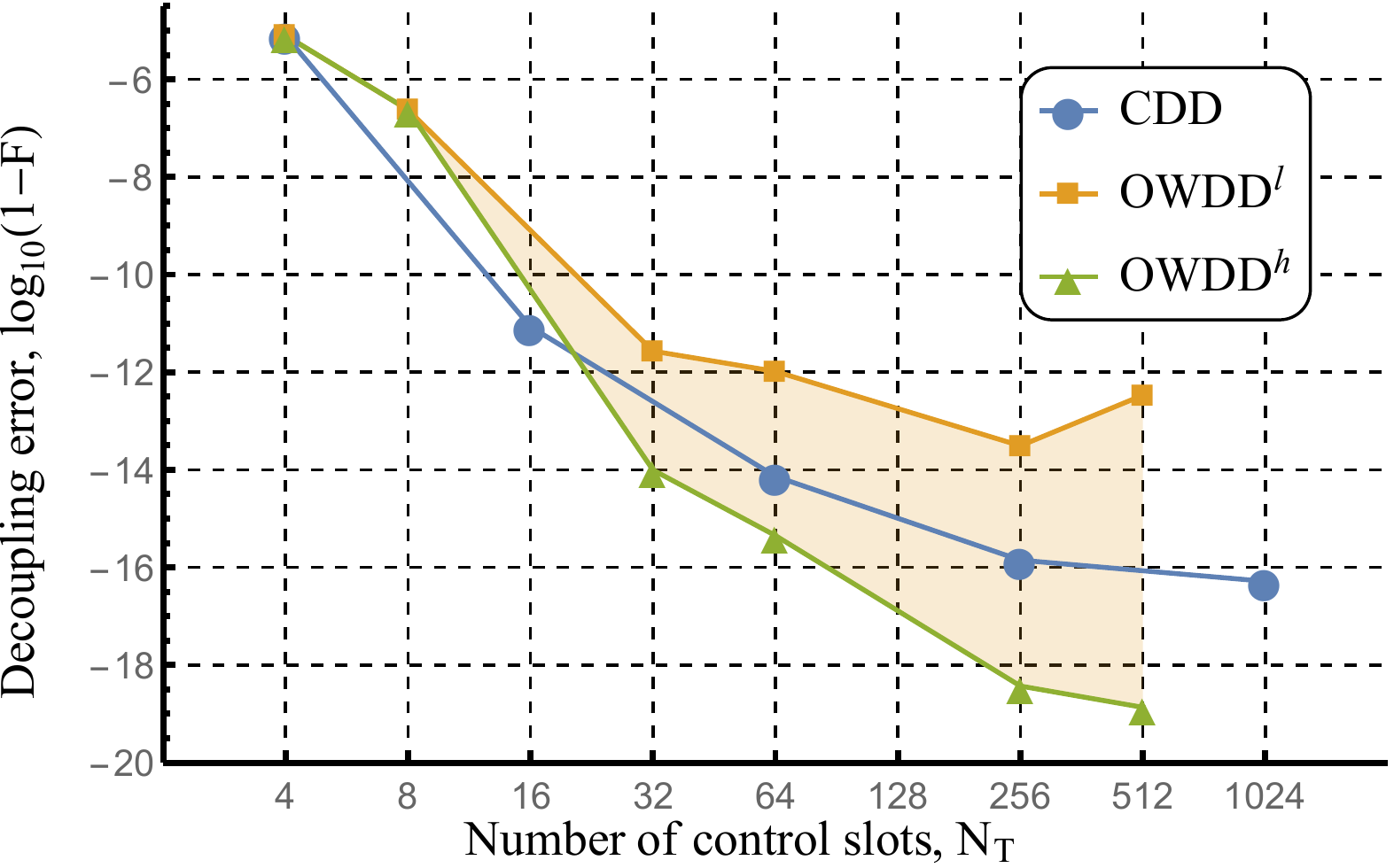}
\vspace*{-1mm}
\caption{(Color online) Fidelity loss vs. $N_T$ for two choices of OWDD$_\alpha$ 
protocols with same CO and CDD. 
The shaded area marks the performance spread expected for all OWDD$_\alpha$ protocols 
in the same equivalence class. A toy model consisting of three 
bath spins is used, with an initial joint state of the form $\ket{\Psi}_{SB} \equiv \ket{\psi} 
\otimes\ket{z_1z_2z_3}$, where $\ket{\psi}$ is a random qubit state and each bath spin is randomly 
chosen over $z_i \in \brac{0,1}$. Results are averaged over 500 realizations. 
We choose parameters $\beta$ = 10kHz, $J=1$MHz, and $\tau_0=0.1\mu$s, suitable for 
qualitatively describing GaAs quantum dots 
\cite{coish2006measurement}.  
}
\label{fig:idealpulse}  
\end{figure}

Since, in each use of Eqs.~(\ref{eq:norm_bound}), the duration of the sequence before 
concatenation enters explicitly, the EPG  
of two GWDD sequences with the same number of projections along each direction
will still differ depending on the order in which the projections are taken ---   
resulting in different fidelities for the same CO. 
To illustrate this sensitivity to the control path, we compare two different choices in the 
same OWDD$_\alpha$ equivalence class with CDD.  Let OWDD$^{{h}}_\alpha \equiv$ 
$\lbrace\text{CPDD}_{xy}$, $\text{CPDD}_{xyz}$, $\text{CPDD}_{xyzxy}$, $\text{CPDD}_{(xyz)^2},...\rbrace$, 
and OWDD$^{{l}}_\alpha \equiv $ $\lbrace\text{CPDD}_{xy}$,$\text{CPDD}_{xyz}$, 
$\text{CPDD}_{xxyyz}$, $\text{CPDD}_{x^2 y^2 z^2},...\rbrace$, for $\alpha=1, 2, 3, 4$. 
Both obey Eqs. (\ref{eq:OWDD-conditions-1}) and (\ref{eq:OWDD-conditions-2}) for the same  
$(r_x,r_y,r_z)$. As detailed in Ref.~\cite{supplement}, OWDD$^{{l \,(h)}}_\alpha$ sequences 
result in comparatively high (low) EPG due to their larger (smaller) prefactor: e.g.,   
at $\text{CO}=3$, one finds $\text{EPG}^{ {\text{OWDD}^h_3}} \leq 5\cdot 2^5 (\tau_0 \beta)^3 J T$ 
vs. $\text{EPG}^{{\text{OWDD}^l_3}} \leq 16\cdot 2^5 (\tau_0\beta)^3 J T$, 
with $T = 2^5\tau_0$, and the difference further increasing for higher CO.  
Geometrically, if one visualizes the implemented sequence of projections in terms of a 
lattice path starting at the origin in $\mathbb{N}^3$, OWDD$^h$ maximizes the number 
of switches in direction as compared to OWDD$^l$. 
That avoiding control path repetitions is generally useful in slowing down 
coherent 
error build-up, has been emphasized in the context of 
randomized DD design \cite{randomDD}, and we conjecture that a similar intuition may  
be key for further optimizing OWDD against path variations. 

We conclude by comparing in Fig. \ref{fig:idealpulse} 
the performance of OWDD and CDD directly in terms of average fidelity loss, 
by resorting to an exact numerical simulation of a low-dimensional spin-bath model, 
which mimics the basic
features of hyperfine-induced decoherence of an electron spin qubit in a quantum dot
\cite{de2003theory,coish2006measurement}. 
The bath operators are $B_\mu = \sum_{i\neq j}\sum_{\alpha,\beta}
c^\mu_{\alpha\beta}(\sigma^\alpha_i\otimes \sigma^\beta_j)$, where $i,j$ index the 
bath qubits, $\mu,\alpha,\beta\in\brac{0,x,y,z}$, and 
$c^\mu_{\alpha\beta}$ are uniformly random coupling constants in $[0,1]$ . 
We assume a fixed minimum pulse interval $\tau_0=0.1\mu$s. 
At large $N_T$, the performance tends to plateau (or even deteriorate) 
due to the fact that convergence breaks down 
for long $T$ (see also Ref.~\cite{prefactor}). 
Remarkably, if OWDD$_\alpha^h$ is used, comparable performance to CDD$_{\alpha}$ is found 
for smaller $N_T$, whereas for same $N_T$, the fidelity of OWDD can be higher than the one 
of CDD by {\em up to two orders of magnitude}.

In a realistic scenario, pulse imperfections are an additional important factor in limiting achievable
operational fidelities. While we leave the study of realistic control errors to a future separate investigation, 
it is worth noting that the robust RGA$_8$ family identified in \cite{quiroz2013optimized} is built 
by suitably incorporating phase alternation into the even orders of OWDD$^h$, pointing to 
an interesting venue for generalization. A characterization of OWDD in terms of 
control symmetry properties (including ``displacement anti-symmetry'' as in 
\cite{njp2016}) and a more rigorous understanding of path sensitivity are also 
well worth pursuing, along with extensions to multi-qubit DD settings.

It is a pleasure to thank Gregory Quiroz, Gerardo Paz-Silva, Kaveh Khodjasteh, 
Leigh Norris, and Manish Gupta for helpful discussions.
LV gratefully acknowledges support from the US Army Research Office under contract No. 
W911NF-14-1-0682. HQ and JPD are supported by the Air Force Office of Scientific Research, 
the Army Research Office, the National Science Foundation and the Northrop Grumman Corporation.


%

\onecolumngrid

\vspace*{5mm}

\begin{center}
	{\large \bfseries Supplementary Material}
\end{center}

\section{Upper bound on EPG for general Walsh DD sequences}

In this section, we provide a detailed derivation of Eq. (5),
and the upper bound to the EPG quoted in the main text for arbitrary GWDD sequences. 
Thanks to the equivalence between GWDD and CPDD, we only need to calculate the upper bound for the corresponding CPDD sequence. The geometrical picture of projections makes CPDD the natural framework to use. Specifically, we first show how the norm of the relevant interaction Hamiltonian is renormalized by a single projection sequence. Since every CPDD sequence arises from concatenation of a series of  projections, we can then apply the result of a single projection recursively, to establish the desired upper bound.

\subsection{Bath renormalization by a single projection}
\label{subsec:bathRenorm}

Consider first the effect of a single projection sequence, say $\mathtt{p_x}$. 
The resulting toggling-frame error Hamiltonian is 
\begin{eqnarray}
\label{eq:piecewise}
\tilde{H}_e(t) = \begin{cases}
H, ~~0\leq t \le \tau_0,\\
XHX, ~~\tau_0\leq t \le 2\tau_0.
\end{cases}
\end{eqnarray}
Since $\tilde{H}_e(t)$ is a piece-wise constant function, the first three orders of the Magnus series 
expansion may be easily computed as 
\begin{eqnarray*}
	\Omega_e^{(1)} &=& \tau_0(H + XH X),\\
	\Omega_e^{(2)} &=& -\frac{i}{2}\tau_0^2[H, XH X],\\
	\Omega_e^{(3)} &=& \frac{1}{3!}\tau_0^3[XH X[H, XH X]].
\end{eqnarray*}
By using the explicit form of $H= \sum_{\mu=0,x,y,z} \sigma_\mu \otimes B_\mu$ given in the main 
text, together with Eq. \eqref{eq:piecewise} above, the first two contributions become 
\begin{eqnarray}
\label{eq:A1} 
\Omega_e^{(1)} &=& 2\tau_0(\unit\otimes B_0 + B_x\otimes X),\\
\label{eq:A2} 
\Omega_e^{(2)} &=& Y\otimes \tau_0^2(i[B_0,B_y] + \brac{B_z,B_x}) 
+ Z\otimes \tau_0^2(i[B_0,B_z] + \brac{B_y,B_x}), 
\end{eqnarray}
with a corresponding norm
\begin{eqnarray*}
	||\Omega_e^{(1)}|| = O(\tau_0(\beta +J)),\quad \quad 
	||\Omega_e^{(2)}|| = O(\tau_0^2J(\beta +J)) .
\end{eqnarray*}
Although the Magnus expansion converges as long as $||H|| T < \pi$, care is needed in discarding 
higher-order terms. The norm of the third-order term is found to be 
\begin{eqnarray}
\label{eq:3rd-order}
||\Omega_e^{(3)}|| = O(\tau_0^3\beta J(\beta +J)) +O(\tau_0^3J^3).
\end{eqnarray}
Accordingly, it is not possible in general to ignore this contribution as it is not clear 
which term in Eq. \eqref{eq:3rd-order} dominates. Following the analysis in 
K. Khodjasteh and D. A. Lidar, Phys. Rev. A {\bf 75}, 062310 (2007), 
we proceed by addressing separately two limiting regimes: 

$\bullet$ When $J \ll \beta$, we have $||\Omega_e^{(1)}|| = 
O(\tau_0\beta)$, $||\Omega_2^{(2)}|| = O(\tau_0\beta J)$ and $||\Omega_e^{(i)}|| = O(\tau_0^i\beta^{i-1}J).$ Therefore, we have
\begin{eqnarray}
\label{eq:Acompare}
||\Omega_e^{(1)}|| < ||\Omega_e^{(2)}|| \ll ||\Omega_e^{(i\geq 3)}||,
\end{eqnarray} 
as long as the condition $\beta \tau_0\ll 1$ is obeyed.


$\bullet$ When $J \gg\beta$, we have  $||\Omega_e^{(i)}|| = O(\tau_0^iJ^i)$. Thus, the same relation 
given in Eq. \eqref{eq:Acompare} holds, as long as $J\tau_0\ll 1$.  

\vspace*{1mm}

In summary, when $J \gg\beta$ or $J\ll\beta$, provided that $\tau_0||H||\ll 1$, it suffices to retain  
the first two orders of the Magnus expansion, giving an approximate expression for the error action 
operator as 
\begin{eqnarray*}
	\Omega_e(2\tau_0) \approx \Omega^{(1)} (2\tau_0) +\Omega^{(2)} (2\tau_0) 
	\equiv 2\tau_0\bar{H}^x = 2\tau_0 \sum_{\mu=0,x,y,z} \sigma_\mu\otimes B^{x}_\mu, 
\end{eqnarray*}
where in the last equality we have defined the average Hamilton associated with $\mathtt{p}_x$ and the relevant renormalized bath operators. 
From Eqs. (\ref{eq:A1}) and  (\ref{eq:A2}), we can read them off as 
\begin{eqnarray*}
	B_0^{x} &=& B_0,  \\\nonumber
	B_x^{x} &= &B_x,\\\nonumber
	B_y^{x} &=& \frac{\tau_0}{2}(i[B_0,B_y] + \brac{B_z,B_x}),\\\nonumber
	B_z^{x} &=& \frac{\tau_{0}}{2}(i[B_0,B_z] + \brac{B_y,B_x}).
\end{eqnarray*}
Similar equations hold for projections along the $y$ or $z$ directions.
When the strength of the system-bath interaction and the pure bath dynamics are of the 
same order of magnitude, $J\sim \beta$, the calculation depends on the specific value of $J$ 
and $\beta$, and no general analytic error bound may be established. 
From now on, we thus assume that the system is in either of the two regimes mentioned above.

\subsection{Bath renormalization in an arbitrary CPDD sequence}

Consider a CPDD sequence specified by an ordered string $s_0$, with total running time $T_{s_0}$. 
Let the relevant effective Hamiltonian be denoted by $\bar{H}^{s_0}$. We now construct a new CPDD 
sequence by concatenating it with a projection sequence, say, $\mathtt{p}_x$, obtaining a 
CPDD$_{s_0x}$, whose renormalized effective Hamiltonian $\bar{H}^{s_0x}$ we wish to determine.

The evolution propagator of the system under the control of CPDD$_{s_0x}$ is 
\begin{eqnarray*}
	X(e^{-i\bar{H}^{s_0}T_{s_0}})X(e^{-i\bar{H}^{s_0}T_{s_0}})  =e^{-iXe^{-i\bar{H}^{s_0}T_{s_0}}X}e^{-ie^{-i\bar{H}^{s_0}T_{s_0}}}.
\end{eqnarray*}
Therefore, the toggling-frame error Hamiltonian 
$\bar{H}^{{s_0}x}(t)$ is still a piece-wise constant function,
\begin{eqnarray*}
	\bar{H}^{{s_0}x}(t) = \begin{cases}
		\bar{H}^{{s_0}}~~, 0\leq t \le T_{s_0},\\
		X\bar{H}^{s_0}X ~~,T_{s_0}\leq t \le 2T_{s_0}, 
	\end{cases}
\end{eqnarray*}
which makes it possible to use the same analysis used in the previous section. 
Accordingly, in the two regimes where $J \ll \beta$ or $J \gg \beta$, the renormalized bath operators are given by
\begin{eqnarray}
B_0^{{s_0}x} &=& B_0^{{s_0}},\nonumber \\
B_x^{{s_0}x} &=& B_x^{{s_0}},\nonumber \\
B_y^{{s_0}x} &=& \frac{T_{s_0}}{2}\left(i[B_0,B_y^{s_0}] + \brac{B_z^{s_0},B_x^{s_0}}\right),\nonumber\\
B_z^{{s_0}x} &=& \frac{T_{s_0}}{2}\left(i[B_0,B_z^{s_0}] + \brac{B_y^{s_0},B_x^{s_0}}\right), 
\label{eq:bathRenorm}
\end{eqnarray}
where $B_\mu^{s_0}, \mu \in\brac{0,x,y,z}$ are the effective bath operators of CPDD$_{s_0}$. As we can see, 
$\mathtt{p}_x$ leaves $B^{s_0}_x$ unchanged, but renormalizes $B^{s_0}_y$ and $B^{s_0}_z$ to the next order. 
Similar renormalization relations  hold for $\mathtt{p}_y$ and $\mathtt{p}_z$. 

Eqs.~(5) in the main text follows from applying standard operator-norm inequalities to the 
renormalized bath operators in Eq. (\ref{eq:bathRenorm}), in particular,
$ ||[A,B]|| \leq 2||A||\,||B||,$ $\||A + B|| \leq ||A|| + ||B||,$ and
$||AB|| \leq ||A||\,||B||.$ 
Along with the definition of the EPG, this yields the desired result for CPDD$_s$,
\[ \text{EPG} \leq T_{s} \sum_{u=x,y,z}||B_u^{s}||. \]

\section{Control path sensitivity}

As remarked in the main text, any permutation of the order of concatenation in building CPDD sequences 
will leave the CO invariant. We expect that pulse sequence with a different control path will give different performance, since the EPG (or fidelity) do not solely depends on the CO. In the context of GWDD,  
control path sensitivity may be understood by comparing the upper bounds of the EPG generated by different control paths. As shown by Eq. (5) in the main text, the EPG of CPDD sequences generated by permutations of a sequence $s$, have the same scaling behavior on $\tau_0$, but produce different prefactors. In this section, we first present a concrete example to demonstrate how the information about the control path is ``encoded'' into the prefactors of the relevant EPG. We then provide a more convenient way to calculate the prefactor for any GWDD/CPDD sequences, whereby we also derive the relevant prefactors for the 
OWDD$^{l\,(h)}$ sequences analyzed in the main text.

\subsection{Switches in the control path are good for error suppression}

The simplest non-trivial example we may consider is to compare CDD$_2 = \text{CPDD}_{xyxy}$ 
with the CPDD sequence generated by a permutation of $xyxy$, denoted by $\overline{\text{CDD}}_2=\text{CPDD}_{xxyy}$. 
To simplify the calculation and to focus on prefactor, we assume the regime 
$\beta \gg J$. Applying the renormalization given in Eq. (5) repeatedly, we have for CDD$_2$
\begin{align*}
||B_x^{xyxy}|| & \leq 2^3\cdot 2^1 \tau_0^2\beta^2 J,\\
||B_y^{xyxy}|| &\leq 2^2\cdot 2^0 \tau_0^2\beta^2 J, 
\end{align*}
where at each step we only keep the leading-order terms. The bound for $||B_z||$ is always higher order than the other two directions since both $\mathtt{p}_x$ and $\mathtt{p}_y$ suppress $B_z$. In the above equations we also see explicitly how the prefactors are accumulated. Similarly, for $\overline{\text{CDD}}_2$, we apply Eq.(5) repeatedly but with a different order, obtaining 
\begin{align*}
||B_x^{xxyy}|| &\leq 2^3\cdot 2^2 \tau_0^2\beta^2 J,\\
||B_y^{xxyy}|| &\leq  2^1\cdot 2^0 \tau_0^2\beta^2 J.
\end{align*}
As we can see, the upper bounds of CDD$_2$ and $\overline{\text{CDD}}_2$ have the same scaling 
over $\tau_0$, consistent with the fact that both of them achieve $\text{CO}=2$. However, CDD$_2$ has a smaller prefactor than $\overline{\text{CDD}}_2$:
\begin{align*}
\text{EPG}^{xyxy} &\leq 20 \, \tau_0^2\beta^2 J T\\
\text{EPG}^{xxyy} &\leq  34 \, \tau_0^2\beta^2 J T.
\end{align*}
This can be qualitatively explained as follows. When $\mathtt{p}_x$ is applied, the upper bound for 
$||B_y||$ will start to accumulate a prefactor. If we continue applying $\mathtt{p}_x$, like in 
$\overline{\text{CDD}}$, the prefactor for $||B_y||$ will grow exponentially since the length of the 
sequence is exponentially increasing. However, if the direction of the projection sequence is changed at a certain point, say to $\mathtt{p}_y$, then the prefactor for $||B_y||$ will stop increasing. Therefore, CPDD sequences, with a large number of switches in the direction of the corresponding projections, 
tend to have lower error and better performance.
For sufficiently large CO we may write  
\begin{align*}
\text{EPG}^{\text{CDD}_\alpha} & \leq 2^{\alpha^2} (\tau_0\beta)^\alpha J T,\\
\text{EPG}^{\overline{\text{CDD}}_\alpha}  & \leq  2^{\frac{1}{2}(3\alpha^2-\alpha)} 
(\tau_0\beta)^\alpha J T.
\end{align*}
The above conclusions remain unchanged if we work in the opposite regime, $\beta \ll J$, since the prefactor only depends on the order of concatenations.

\subsection{Calculating prefactors for GWDD/CPDD sequences}

The method we described above to calculate the prefactors for GWDD sequences relies upon the geometric picture of CPDD. However, the calculation is tedious, especially for long pulse sequences. Here, we present an alternative method to directly calculate the prefactor for any GWDD/CPDD sequence.

Consider a pulse sequence CPDD$_s$. Then:
\begin{enumerate}
	\item 
	Define $s'$ to be the sequence of letters in the reverse order of $s$, namely,  $s' \equiv s_1\ldots s_m$.
	Construct a $3\times |s|$ matrix, denoted by $\mathcal{L}$, according to the following rule:
	\begin{align}
	\label{def:L-matrix}
	\mathcal{L}_{\mu j} \equiv \begin{cases}
	1, ~\text{if}~ s_j = \mu~\\
	0, ~\text{otherwise}
	\end{cases}~.
	\end{align}
	where we use $\mu \in \brac{x,y,z}$ to label the $1^{\text{st}}$, the $2^{\text{nd}}$ and the $3^{\text{rd}}$ row of $\mathcal{L}$.
	\item The prefactor in the upper bound on $||B_\mu^s||$ is then given by
	\begin{align}
	\label{eq:construct-prefactor}
	\prod_{\substack{j=1\\
			\bar{\mathcal{L}}_{\mu j}\neq 0}}^{|s|} \bar{\mathcal{L}}_{\mu j} 2^{j-1}~,
	\end{align}
	where the matrix $\bar{\mathcal{L}}$ is the logical negation of $\mathcal{L}$.	
	\item  If we assume $\beta\gg J$ and ignore higher-order contributions, we have the following upper bound
	\begin{align}
	\label{eq:construct-upper-bound}
	||B_\mu^s||\leq \left(\prod_{\substack{j=1\\
			\bar{\mathcal{L}}_{\mu j}\neq 0}}^{|s|} \bar{\mathcal{L}}_{\mu j} 2^{j-1}\right) (\tau_0\beta)^{\sum_{k=1}^{|s|}\bar{\mathcal{L}}_{\mu k}} J~.
	\end{align}
\end{enumerate}

We illustrate the above procedure by considering a simple example, namely, the second level of OWDD sequences, CPDD$_{xyz}$. From the definition of CPDD sequence, $s'=xyz$, hence the matrix $\mathcal{L}$ is given by
\[
\begin{blockarray}{cccccc}
s'&&& x & y & z\\
\begin{block}{ccc[ccc]}
x&&& 1 &0&0\\
y&&& 0 &1&0\\
z&&& 0 &0&1\\	
\end{block}
\end{blockarray}~.
\]
Here, the row indexes represent different directions while the column indexes are specified by $s'$. Applying Eq. \eqref{eq:construct-prefactor} and Eq. \eqref{eq:construct-upper-bound}, and assuming $\beta\gg J$, we get
\begin{align*}
||B_x^{xyz}||\leq 2^2\cdot 2^1 (\tau_0 \beta)^2 J,\\
||B_y^{xyz}||\leq 2^2\cdot 2^0 (\tau_0\beta)^2 J,\\
||B_z^{xyz}||\leq 2^1\cdot 2^0 (\tau_0\beta)^2 J.
\end{align*} 

\subsection{Path sensitivity for optimal Walsh DD sequences}

Any GWDD sequence that achieves $\text{CO}= \alpha$, with a number of time slots 
obeying $\log_2(N_T) = \frac{1}{2}\left( 3\alpha+ \overline{\alpha}  \right)$, as explained in the main 
text [see also Table II], is an OWDD$_\alpha$ sequence.  
Although different choices of OWDD use the same number of control time slots for given CO, their performance is different due to the control path sensitivity discussed above. 
Based on the intuitive argument we described, we expect that 
OWDD sequences with a larger number of switches will 
comparatively achieve a lower EPG, hence higher fidelity. With this intuition in mind, we consider 
OWDD sequences with the maximum number of switches, 
$$ \text{OWDD}^{h}_\alpha \equiv \brac{\text{CPDD}_{xy},\text{CPDD}_{xyz}, \text{CPDD}_{xyzxy}, \text{CPDD}_{(xyz)^2},\ldots},$$ 
as well as sequences where this number is minimized and the lattice-path trajectory has 
long straight segments: 
$$\text{OWDD}^{l}_\alpha \equiv \text{CPDD}_{ x^{r_x}y^{r_y}z^{r_z} }  =  
\brac{\text{CPDD}_{xy}, \text{CPDD}_{xyz}, \text{CPDD}_{xxyyz}, \text{CPDD}_{xxyyzz}, \ldots }  , $$ 
as also defined in the main text. 

The first two orders of OWDD sequences are the same for any choice of OWDD. Thus, in order to illustrate the control path sensitivity, below we explicitly calculate the upper bound of EPG for the next two levels of OWDD$^{h}$ and OWDD$^{l}$, corresponding to $\text{CO}=3, 4$, respectively.  

To calculate the upper bound for OWDD$^{h}_3$ = CPDD$_{xyzxy}$, we first write down the $\mathcal{L}$ matrix according to Eq.\eqref{def:L-matrix},
\begin{align}
\mathcal{L}=\begin{bmatrix}
1&0&0&1&0\\
0&1&0&0&1\\
0&0&1&0&0
\end{bmatrix}~.
\end{align}
Calculate its negation $\bar{\mathcal{L}}$ and then the upper bound according to Eq.\eqref{eq:construct-upper-bound}:
\begin{align*}
||B_x^{xyzxy}||&\leq 2^4\cdot 2^2 \cdot 2^1 (\tau_0\beta)^3 J,\\
||B_y^{xyzxy}||&\leq 2^3\cdot 2^2 \cdot 2^0 (\tau_0\beta)^3 J,\\
||B_z^{xyzxy}||&\leq 2^4\cdot 2^3 \cdot 2^1 \cdot 2^0 (\tau_0 \beta)^4 J.
\end{align*}
By discarding the higher-order contribution from $||B_z||$, we have
$\text{EPG}^{\text{OWDD}^{h}_3} \leq 5\cdot 2^5 (\tau_0\beta)^3 JT$, where $T = 2^5\tau_0$. 
Following the same procedure for OWDD$^{h}_4$, we have
\begin{align*}
||B_x^{xyzxyz}||&\leq 2^5\cdot 2^4\cdot 2^2 \cdot 2^1 (\tau_0 \beta)^4 J,\\
||B_y^{xyzxyz}||&\leq 2^5\cdot 2^3\cdot 2^2 \cdot 2^0 (\tau_0 \beta)^4 J,\\
||B_z^{xyzxyz}||&\leq 2^4\cdot 2^3 \cdot 2^1 \cdot 2^0 (\tau_0\beta)^4 J,
\end{align*}
the EPG is dominated by the $x$ direction and we have $\text{EPG}^{\text{OWDD}^{h}_4} \leq 2^{12} \, (\tau_0 \beta)^4 JT$, 
with $T=2^6\tau_0$. 

Now we calculate the upper bounds for OWDD$^l_\alpha$. To calculate the upper bound of OWDD$^{l}_3$, we write down the $\mathcal{L}$ matrix first, namely, 
\begin{align}
\mathcal{L}=\begin{bmatrix}
1&1&0&0&0\\
0&0&1&1&0\\
0&0&0&0&1
\end{bmatrix}~.
\end{align}
Then the upper bounds of bath operators are given by Eq. \eqref{eq:construct-upper-bound},
\begin{align*}
||B_x^{xxyyz}||&\leq 2^4\cdot 2^3 \cdot 2^2 (\tau_0 \beta)^3 J,\\
||B_y^{xxyyz}||&\leq 2^4\cdot 2^1 \cdot 2^0 (\tau_0 \beta)^3 J,\\
||B_z^{xxyyz}||&\leq 2^3\cdot 2^2\cdot 2^1\cdot 2^0 (\tau_0 \beta)^4 J.
\end{align*}
Therefore, $\text{EPG}^{\text{OWDD}^{l}_3} \leq 2^4\cdot 2^5 (\tau_0\beta)^3 JT$, 
with $T=2^5 \tau_0$.
Similarly, the upper bounds for OWDD$_4^l$ are given by
\begin{align*}
||B_x^{xxyyzz}||&\leq 2^5\cdot 2^4\cdot 2^3 \cdot 2^2 (\tau_0\beta)^4 J,\\
||B_y^{xxyyzz}||&\leq 2^5\cdot 2^4\cdot 2^1\cdot 2^0 (\tau_0 \beta)^4 J,\\
||B_z^{xxyyzz}||&\leq 2^3\cdot 2^2\cdot 2^1\cdot 2^0 (\tau_0\beta)^4 J,
\end{align*}
whereby $\text{EPG}^{\text{OWDD}^{l}_4} \leq  2^{14}\, (\tau_0\beta)^4 JT$, with $T=2^6\tau_0$.

As one can see from the above calculations, at $\text{CO}=3$ the EPG upper bound of 
OWDD$^{h}$ is about $3.2$ times smaller than the one for OWDD$^{l}$, and 
becomes four times smaller  at $\text{CO}=4$. 
Therefore, an increasingly larger benefit is expected also in terms of fidelity from using OWDD$^{h}$ 
with larger CO.
\end{document}